\documentclass[twoside,12pt]{article}
\usepackage{epsfig}

\newcommand{\be}{\begin{equation}}
\newcommand{\ee}{\end{equation}}
\newcommand{\bea}{\begin{eqnarray}}
\newcommand{\eea}{\end{eqnarray}}

\begin{document}

\title{ \vspace{1cm} Resonance-Parton Duality in $(e,e'\pi)$ off Nucleons
}
\author{Murat Kaskulov and Ulrich Mosel
\\
Institut fuer Theoretische Physik, Universitaet Giessen, D-35392 Giessen, Germany\\
}
\maketitle
\begin{abstract}
Regge-pole based descriptions of pion-electroproduction on
nucleons have given a very good description of the longitudinal components of
the cross sections. However, these very same models grossly underestimate
the transverse components. A related problem appears in QCD-based scaling arguments that predict the predominance of longitudinal over transverse
electroproduction of pions by terms $\propto Q^2$. However, data from JLAB,
Cornell and DESY, covering a wide kinematical range $1 < Q^2 < 11\, {\rm
  GeV}^2$ and $2\, {\rm GeV} < W < 4\, {\rm GeV}$, do not show this expected
behavior. We address here
this issue of the transverse response in pion-electroproduction
by considering the contributions of high-lying ($W > 2$
GeV) nucleon resonances to pion production. The coupling strengths and form
factors are obtained through resonance-parton duality. We show that
in a wide range of electron energies and four-momentum
transfers such a model describes all the available data very well.
\end{abstract}
\section{Introduction}
Electroproduction of mesons in the deep inelastic scattering (DIS),
is a modern tool which permits to study the structure of the nucleon on the partonic
level. Exclusive channels in DIS are of particular importance.
For instance, arguments based on asymptotic QCD predict that the transverse component of the cross section for the exclusive reaction $p(e,e'\pi^+)n$ falls off with $1/Q^8$ while the
longitudinal component falls with $1/Q^6$
\cite{Brodsky:1974vy,Collins:1996fb}. Exclusive pion production can be used to
check this prediction and to determine in which kinematical regime these
simple scaling laws become effective. In the following we discuss these
reactions; we draw here on our publications
\cite{Kaskulov:2008xc,Kaskulov:2009gp,Kaskulov:2010kf} where further details
can be found.

At Jefferson Laboratory (JLAB) the exclusive
reaction $p(e,e'\pi^+)n$  has been investigated for a range of photon
virtualities up to $Q^2\simeq 5$~GeV$^2$ at an invariant mass of the $\pi^+n$
system around the onset of deep--inelastic regime,
$W\simeq 2$~GeV~\cite{Horn:2006tm,Horn:2007ug,Tadevosyan:2007yd}.
A separation of the cross section into the transverse $\sigma_{\rm T}$
and longitudinal $\sigma_{\rm L}$ components has been performed.
The data  show that $\sigma_{\rm T}$ is large at JLAB
energies~\cite{Horn:2007ug}. At $Q^2 = 3.91$~GeV$^2$ $\sigma_{\rm T}$
is by about a factor of two larger than $\sigma_{\rm L}$, contrary to the
QCD-based expectations, and at
$Q^2=2.15$~GeV$^2$ it has same size as $\sigma_{\rm L}$. Previous measurements
at values of $Q^2=1.6~(2.45)$~GeV$^2$~\cite{Horn:2006tm} show a similar
problem in the understanding of $\sigma_{\rm T}$. Even at smaller
JLAB~\cite{Tadevosyan:2007yd} and much higher Cornell~\cite{Bebek:1977pe} and
DESY~\cite{:2007an} values of $Q^2$ there is a disagreement between the
scaling expectations and experimental data.

The longitudinal cross section $\sigma_{\rm L}$ is well understood in terms of
the pion quasi--elastic knockout mechanism~\cite{Neudatchin:2004pu} because of
the pion pole at low $-t$ . This makes it possible to study the charge form
factor of the pion at momentum transfer much bigger than in the scattering of
pions from atomic electrons~\cite{Sullivan:1970yq}. However, the model of
Ref.~\cite{Vanderhaeghen:1997ts}, which is based on such a picture and which
is generally considered to be a guideline for the experimental analysis and
extraction of the pion form factor, underestimates largely the trasnverse
response $\sigma_{\rm T}$ at high values of $Q^2$~\cite{Horn:2007ug}.

\section{Transverse strength in electroproduction}
\begin{figure*}[h]
\begin{center}
\includegraphics[clip=true,width=0.95\columnwidth,angle=0.]{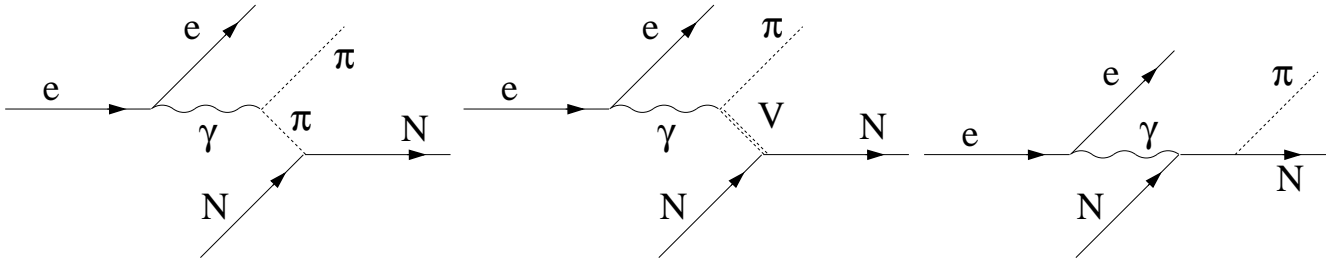}
\caption{\small
 The diagrams describing
       the hadronic part of the $\pi^+$-- electroproduction amplitude at high
       energies. The leftmost diagram shows the $t$-channel contribution, the
       middle one the contribution of vector meson exchange and the rightmost
       one gives the nucleon Born term.   \vspace{-0cm}   } \label{Figure1}
\end{center}
\end{figure*}

The model of Ref.\ \cite{Vanderhaeghen:1997ts} can be represented by the three amplitudes shown in Fig.\ \ref{Figure1}.
The leftmost Reggeized $t$-channel term dominates the longitudinal strength
which thus always dominates at forward momenta, with relatively small
contributions from vector meson exchange (middle graph). The right graph,
involving a nucleon-pole $s$-channel diagram, is necessary for gauge
invariance; its main, but small, contribution is to the transverse cross
section. At the kinematics of the relevant electroproduction experiments
the energies are high enough to excite also nucleon resonances; the
invariant masses of the $(\gamma^*p)$ system are between approximately 2
and 4 GeV. We identify these high-lying resonances with partonic
excitations leading to DIS \cite{Domokos:1971ds} and invoke the correspondence
principle in going from an inclusive final DIS state to the exclusive pion
production~\cite{Bjorken:1973gc}. During this transition the transverse
strength of DIS remains intact. We thus expect that the inclusion of these
resonance excitations into the $s$-channel diagram enhances the transverse
scattering while leaving the longitudinal strength originating in the
$t$-channel diagram intact.
The resonance excitations are also
$s$-channel contributions which -- because of their special coupling --
are gauge invariant by themselves \cite{Penner:2002md}.
Adding such a resonance (DIS) contributions to
the Born term in the model of \cite{Vanderhaeghen:1997ts}
constitutes the central point of our model which also contains an
improved treatment of gauge invariance accounting for the difference
in the electromagnetic form factors for pions and
proton~\cite{Kaskulov:2008xc,Kaskulov:2009gp,Kaskulov:2010kf}.

\begin{figure}[ht]
\begin{center}
\includegraphics[clip=true,width=0.50\columnwidth,angle=0.]
{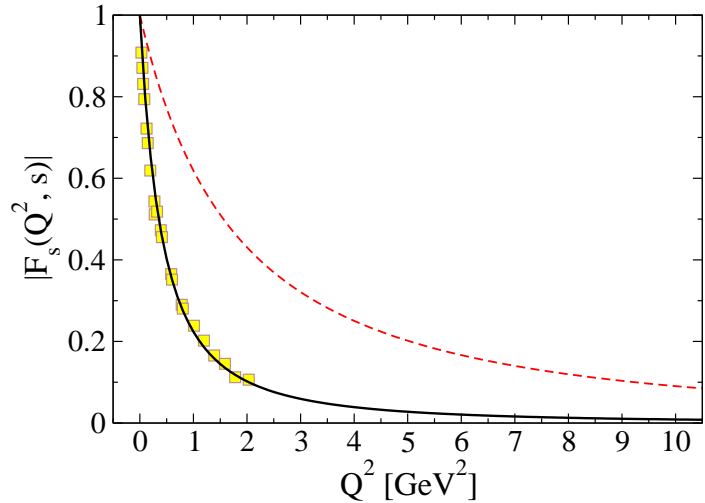}
\caption{
\label{F1onoff} \small
The $Q^2$ dependence of the absolute value
of the transition form factor $|F_s(Q^2,s)|$ in Eq.~\ref{sumres}
(dashed curve) at $\sqrt{s}=2.2$~GeV. The solid curve
describes the proton Dirac form factor in
comparison with data. From~\cite{Kaskulov:2010kf}.
}
\end{center}
\end{figure}

\begin{figure}[h]
\begin{center}
\includegraphics[clip=true,width=0.65\columnwidth,angle=0.]{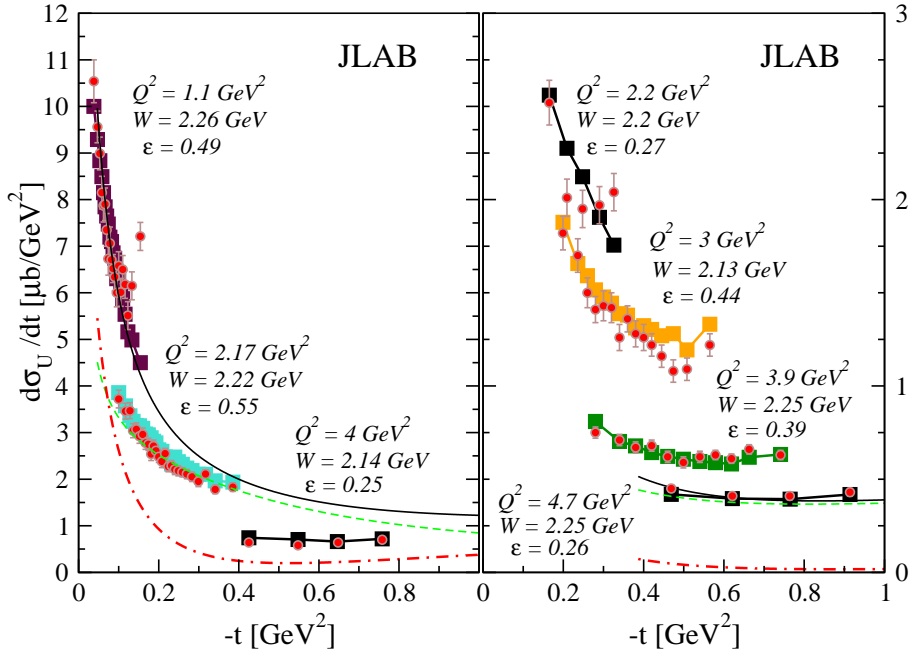}
\caption{\label{Dutta1} \small The differential cross sections
$d\sigma_{\rm U}/dt = d\sigma_{\rm T}/dt + \varepsilon d\sigma_{\rm
L}/dt$ in exclusive reaction $p(\gamma^*,\pi^+)n$
in the kinematics of the $\pi$-CT experiment at JLAB~\cite{:2009ub}.
The square symbols connected by solid lines describe the
model results. The discontinuities in the curves result from
the different values of $(Q^2,W,\varepsilon)$ for the various $-t$ bins.
The dash-dotted and dashed curves describe the contributions of the longitudinal
$\varepsilon d\sigma_{\rm L}$ and transverse $d\sigma_{\rm T}$
cross sections, respectively, to the total unseparated cross sections (solid curves)
for the the lowest and highest average values of $Q^2=1.1$~GeV$^2$ and $Q^2=4.7$~GeV$^2$.}
\vspace{-0.cm}
\end{center}
\end{figure}

\begin{figure}[ht]
\includegraphics[clip=true,width=0.99\columnwidth,angle=0.]
{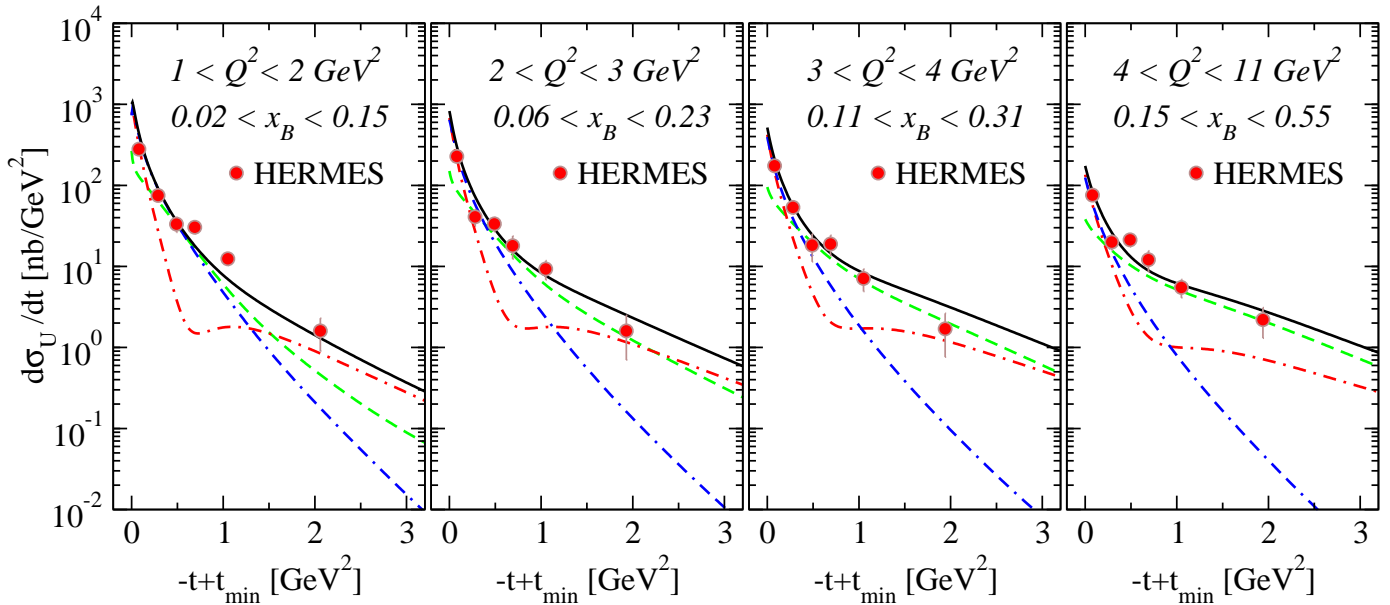}
\caption{\label{EffHermes} \small $-t+t_{min}$ dependence of the
differential cross section $d\sigma_U/dt = d\sigma_{\rm T}/dt + \epsilon d\sigma_{\rm
L}/dt$ in exclusive reaction $p(\gamma^*,\pi^+)n$ at HERMES.
The experimental data are from Ref.~\cite{:2007an}.
The calculations are performed for the average values of
$(Q^2,x_{\rm B})$ in a given $Q^2$ and Bjorken $x_{\rm B}$ bin.
The solid curves are the full model results.
The dash-dotted curves correspond to the longitudinal $\epsilon d\sigma_{\rm L}/dt$
and the dashed curves to the transverse $d\sigma_{\rm T}/dt$ components of the
cross section.
The dash-dash-dotted curves describe the results without the
resonance/partonic effects. From \cite{Kaskulov:2010kf}.}
\vspace{-0.cm}
\end{figure}

\begin{figure}[h]
\begin{center}
\includegraphics[clip=true,width=0.85\columnwidth,angle=0.]{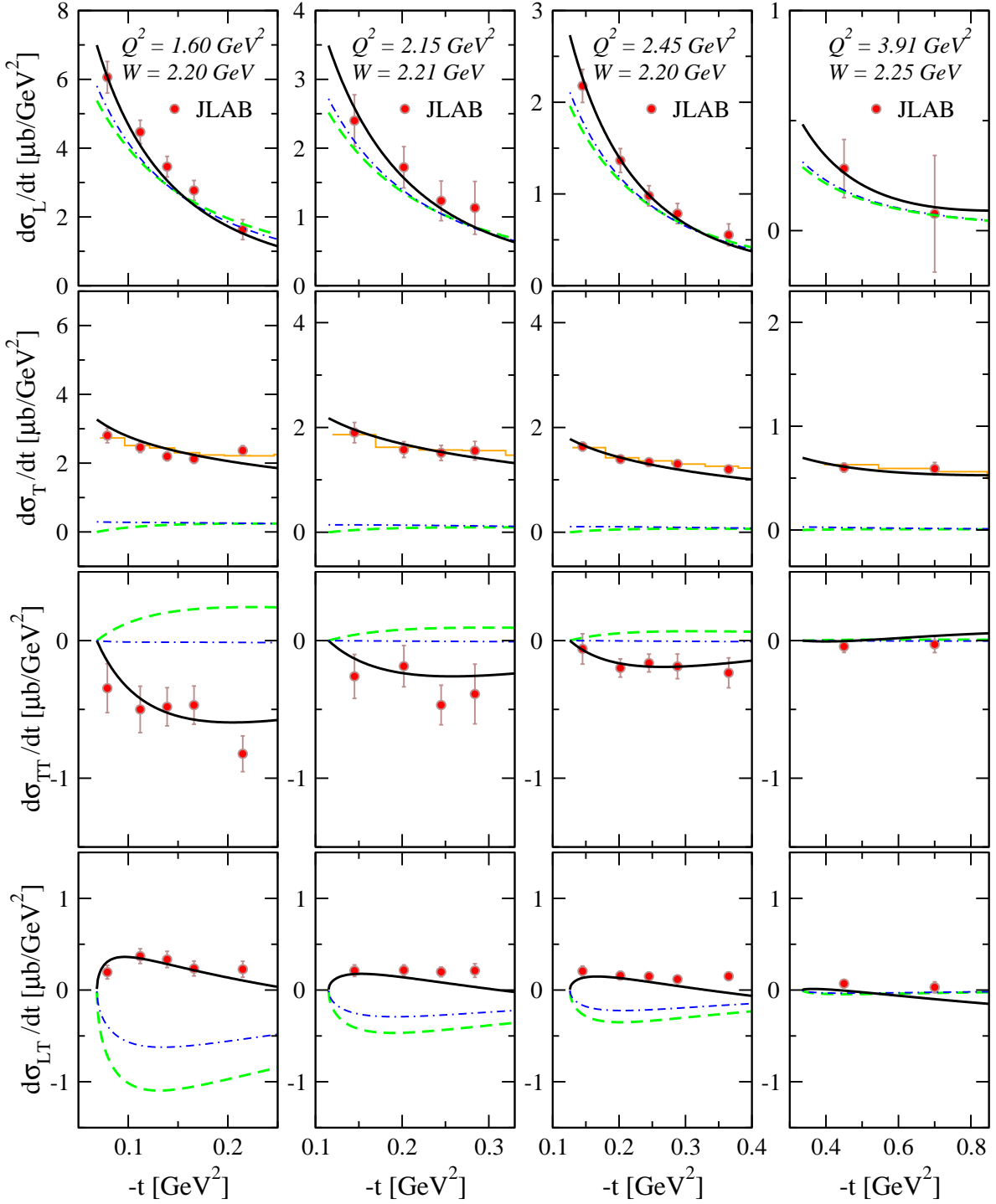}
\caption{\label{Horn1} \small
$-t$ dependence of \textsc{l/t} partial transverse
$d\sigma_{\rm T}/dt$, longitudinal $d\sigma_{\rm L}/dt$ and interference
$d\sigma_{\rm TT}/dt$ and $d\sigma_{\rm LT}/dt$ differential cross sections
in exclusive reaction $p(\gamma^*,\pi^+)n$. The experimental data are from
the $F\pi$-2~\cite{Horn:2006tm} and $\pi$-CT~\cite{Horn:2007ug}
experiments at JLAB. The numbers displayed in the plots are the average $(Q^2,W)$ values.
The dashed curves correspond to the exchange of the $\pi$-Regge trajectory alone.
The dash-dotted curves are obtained with the on-mass-shell form
factors in the nucleon-pole contribution and exchange of the
$\rho(770)/a_2(1320)$-trajectory. The solid curves describe the model results with the
resonance contributions. The data points in each $(Q^2,W)$ bin correspond
to slightly different values of $Q^2$ and $W$ for the various $-t$ bins.
The calculations are performed for values of $Q^2$ and $W$ corresponding to
the first $-t$ bin. The histograms for $d\sigma_{\rm T}/dt$ are the results
from~\cite{Kaskulov:2008xc}. From \cite{Kaskulov:2010kf}.
}
\end{center}
\end{figure}


In the experiments quoted the invariant masses $W$ of the $(\gamma^*,N)$ system
are all $W > 2$ GeV, i.e.\ they all lie above the region of well-established,
separated nucleon resonances. Thus, the Born term, that has become a sum over
individual resonances, can be replaced by an integral over resonances
with average coupling constants and form factors
\begin{eqnarray}      \label{sumres}
B = \sum \limits_{i} r(M_i) c(M_i)
\frac{F(Q^2,M_i^2)}{s-M_i^2+i0^+} &\Rightarrow&
\int \limits_{M_p^2}^{\infty} dM_i^2 \rho(M_i^2) r(M_i^2)c(M_i^2)
\frac{F(Q^2,M_i^2)}{s-M_i^2+i0^+} ~,
\end{eqnarray}
where $r(M_i)$ and $c(M_i)$ are the electromagnetic and strong couplings,
respectively, and $F(Q^2,M_i^2)$ is the electromagnetic form factor.
Here $\rho(M_i^2)$ is the density of resonances with mass $M_i$.
So far unknown are here the couplings $r$ and $c$ and the form
factors $F$ in Eq.\ (\ref{sumres}).

Our aim is to maintain the transverse character of DIS, which follows from a
parton picture, when going to the exclusive limit of a resonance decay.
We thus have to establish a connection between these two pictures.
Bloom and Gilman \cite{Bloom:1970xb,Bloom:1971ye} (BG) have shown that
the total DIS strength follows closely the average behavior of
nucleon resonances (for a more recent review see also
\cite{Melnitchouk:2005zr}). We, therefore, now use this BG
duality in its local form
\begin{equation}
\label{BG}
F_2^p(x_{\rm B},Q^2) = \sum_{i} (M_i^2-M_p^2+Q^2)  W(Q^2,M_i) \delta(s - M_i^2),
\end{equation}
where $x_{\rm B}$ stands for the Bjorken scaling variable and the deep inelastic
structure function $F_2^p(x_{\rm B},Q^2)$ is expressed as a sum of resonances.
$W(Q^2,M_i)$ defines the $i$th resonance contribution to the $\gamma^* p$ forward
scattering amplitude; it is essentially the electromagnetic coupling constant
$r(M_i)$ times the resonance form factor $F(Q^2,M_i)$.
Eq.\ (\ref{BG}) links the partonic content of nucleon resonances with their
hadronic structure. Since the density of resonances in Eq.\ (\ref{sumres})
is a steeply increasing function of invariant mass, it follows from
Eq.\ (\ref{BG}) that the electromagnetic coupling to nucleon resonances
must be decreasing with mass since $F_2$ is finite. This leads to a natural
cut-off for the number of resonances and makes the integral in
Eq.~(\ref{sumres}) finite.

In a further step the combined nucleon-resonance contribution to the s-channel
is absorbed into an effective Born-term with a nucleon pole only, but a
modified form factor $F_s$ which now contains all the effects of the resonances
\begin{equation}
B = \frac{F_s(Q^2,s)}{s-M_p^ 2+i0^+} ~.
\end{equation}
Parameterizing the form factor in a dipole form leads to the conclusion that
the cut-off must increase with the mass of the
resonance, a result that is well known from early studies of BG duality ~\cite{Elitzur:1971tg}. We, therefore, expect that the effective form
factor $F_s$ is considerably harder than that of the nucleon alone.
This is indeed born out by the calculations as can be seen in Fig.\ \ref{F1onoff}.

In Fig.~\ref{Dutta1} we compare the result of our calculations
with the $p(\gamma^*,\pi^+)n$  data from JLAB~\cite{:2009ub}
for unseparated cross sections $d\sigma_{\rm U}/dt$,
at values of $W \simeq 2.2\div 2.4$~GeV and for different values of
$(Q^2,\varepsilon)$ bins. The square symbols connected by solid lines describe the
model results. The discontinuities in the curves result from
the different values of $(Q^2,W,\varepsilon)$ for the various $-t$ bins.
The data are very well reproduced by the
present model in the measured $Q^2$ range from $Q^2 \simeq 1$~GeV$^2$
up to $5$~GeV$^2$.
In Fig.~\ref{Dutta1}
we also show the contributions of the longitudinal
$\varepsilon d\sigma_{\rm L}$ (dash-dotted curves) and transverse $d\sigma_{\rm T}$
(dashed curves) cross sections to the total unseparated cross sections (solid curves)
for the lowest and highest average values of $Q^2=1.1$~GeV$^2$ and $Q^2=4.7$~GeV$^2$.
The cross sections at high values of $Q^2$ are flat and
totally transverse. At forward angles a strong peaking of the cross
section at $Q^2=1.1$~GeV$^2$ comes from the large
longitudinal component in this case. The off-forward region is
transverse. This behavior agrees with the results from~\cite{Kaskulov:2008xc}.
As we shall see, the same behavior is observed in the DIS
regime at HERMES~\cite{:2007an} where the value of $W$ is
higher. At HERMES, because of the Regge shrinkage
of the $\pi$-reggeon exchange and smaller transverse component, the forward peak
just has a steeper $-t$-dependence~\cite{Kaskulov:2009gp}.

In Fig.\ \ref{EffHermes} we show that, indeed, this very same picture also works
remarkably well at the much higher momentum transfers $Q^2$ and invariant
masses $W$ reached in the HERMES experiment \cite{:2007an}. Even in the
kinematical windows $4 < Q^2 < 11$ GeV$^2$ the difference between the
dash-dash-dotted curve (no resonance contributions) and the dashed
curve (resonances included) shows the dramatic impact of the resonance
contributions which are essential in describing the cross section at larger
$-t$. Just contrary to the situation in the JLAB experiment the
longitudinal cross section at HERMES determines the total differential
cross section at small $-t$.

As the HERMES kinematics are quite close to those expected for JLAB at 12 GeV
we predict that again high-lying resonances determine the transverse cross
sections at larger $-t$. In Fig.\ 23 in Ref.\ \cite{Kaskulov:2010kf} detailed
predictions for the $L/T$ separated cross sections are given both for $\pi^+$
and $\pi^-$ production. In particular we predict that $\pi^-$ production is
largely longitudinal.

Fig.~\ref{Horn1} shows the results of the calculations in comparison
with the data of Ref.\ \cite{Horn:2006tm,Horn:2007ug} for all four
separated cross sections. The solid lines that represent the results of our
calculations describe all cross sections, the longitudinal, the
transverse and the interference ones, very well. The transverse
strength is nearly entirely built up by the resonance contributions
that are considerably bigger and fall off much more weakly with
$Q^2$ than the ones obtained from the nucleon Born term alone
(see Fig.~\ref{F1onoff}).  The resonances also contribute about
30\% to the longitudinal cross section at forward angles (small $-t$)
where the major contribution comes from the $t$-channel graph
with the nucleon Born term alone. For the interference cross
sections $\sigma_{\rm TT}$ and $\sigma_{\rm LT}$ the sign even
changes when the resonances are taken into account.

This model has recently been extended  to the photo- and electroproduction of
$\pi^0$ \cite{Kaskulov:2011wd}; again a very good description of all available
data is reached without any new parameters. The transition from
photoproduction ($Q^2 = 0$) to electroproduction shows that the resonance
contributions assume a larger and larger role with increasing $Q^2$. While
at the photon point they just fill in the diffractive dip in the otherwise
Regge-dominated cross section at higher $Q^2$ they become more and more dominant.

\section{Summary}
The large transverse strength observed in various experiments on exclusive
pion production has been a long-standing puzzle. We have resolved this open
problem by adding to the usual $t$\emph{-channel + Born} term description the
contribution of high-lying nucleon resonances as an effective description
of DIS excitations. By using quark-hadron duality we have been able to link
the properties of these resonances to the partonic degrees of freedom. We
thus treat this contribution to the exclusive production as the limiting
case of inclusive DIS processes~\cite{Kaskulov:2008xc,Kaskulov:2009gp}
that are predominantly transverse. Such a model, that makes use only
of quite general average properties of nucleon resonances, is able to
describe all the available exclusive electroproduction data for pions.
In particular the transverse response of nucleons is determined by
their resonance/parton excitations. A discussion of the relevance of the present results for color transparency measurements in exclusive electroproduction of pions
off nuclear can be found in~\cite{Kaskulov:2008ej,Kaskulov:2011hr}.
\clearpage
\section*{Acknowledgments}
This work has been supported by DFG through the TR16 and by BMBF.


\begin{thebibliography}{99}

\bibitem{Brodsky:1974vy}
  S.~J.~Brodsky, G.~R.~Farrar,
  Phys.\ Rev.\  {\bf D11}, 1309 (1975).

\bibitem{Collins:1996fb}
  J.~C.~Collins, L.~Frankfurt and M.~Strikman,
  Phys.\ Rev.\  D {\bf 56}, 2982 (1997).


\bibitem{Kaskulov:2008xc}
  M.~M.~Kaskulov, K.~Gallmeister, U.~Mosel,
  Phys.\ Rev.\  {\bf D78}, 114022 (2008).

\bibitem{Kaskulov:2009gp}
  M.~M.~Kaskulov, U.~Mosel,
  Phys.\ Rev.\  {\bf C80}, 028202 (2009).

\bibitem{Kaskulov:2010kf}
  M.~M.~Kaskulov, U.~Mosel,
  Phys.\ Rev.\  {\bf C81}, 045202 (2010).


\bibitem{Horn:2006tm}
  T.~Horn {\it et al.}, 
  Phys.\ Rev.\ Lett.\  {\bf 97}, 192001 (2006).

\bibitem{Horn:2007ug}
  T.~Horn {\it et al.},
  Phys.\ Rev.\  C {\bf 78}, 058201 (2008).


\bibitem{Tadevosyan:2007yd}
  V.~Tadevosyan {\it et al.}, 
  Phys.\ Rev.\  C {\bf 75}, 055205 (2007).


\bibitem{Bebek:1977pe}
  C.~J.~Bebek {\it et al.}, 
  Phys.\ Rev.\  D {\bf 17}, 1693 (1978).

\bibitem{:2007an}
  A.~Airapetian {\it et al.} [ HERMES Collaboration ],
  Phys.\ Lett.\  {\bf B659}, 486-492 (2008).


\bibitem{Neudatchin:2004pu}
  V.~G.~Neudatchin {\it et al.}, 
  Nucl.\ Phys.\  A {\bf 739}, 124 (2004).

\bibitem{Sullivan:1970yq}
  J.~D.~Sullivan,
  Phys.\ Lett.\  B {\bf 33}, 179 (1970).


\bibitem{Vanderhaeghen:1997ts}
  M.~Vanderhaeghen, M.~Guidal and J.~M.~Laget,
  Phys.\ Rev.\  C {\bf 57}, 1454 (1998);
  M.~Guidal, J.~M.~Laget and M.~Vanderhaeghen,
  Nucl.\ Phys.\  A {\bf 627}, 645 (1997).




\bibitem{Domokos:1971ds}
  G.~Domokos, S.~Kovesi-Domokos and E.~Schonberg,
  Phys.\ Rev.\  D {\bf 3}, 1184 (1971).

\bibitem{Bjorken:1973gc}
  J.~D.~Bjorken, J.~B.~Kogut,
  Phys.\ Rev.\  {\bf D8}, 1341 (1973).

\bibitem{Penner:2002md}
  G.~Penner, U.~Mosel,
  Phys.\ Rev.\  {\bf C66}, 055212 (2002).



\bibitem{Bloom:1970xb}
  E.~D.~Bloom and F.~J.~Gilman,
  Phys.\ Rev.\ Lett.\  {\bf 25}, 1140 (1970).

\bibitem{Bloom:1971ye}
  E.~D.~Bloom and F.~Gilman,
  Phys.\ Rev.\  D {\bf 4}, 2901 (1971).



\bibitem{Melnitchouk:2005zr}
  W.~Melnitchouk, R.~Ent, C.~Keppel,
  Phys.\ Rept.\  {\bf 406}, 127-301 (2005).




\bibitem{Elitzur:1971tg}
  M.~Elitzur,
  Phys.\ Rev.\  D {\bf 3}, 2166 (1971).


\bibitem{:2009ub}
  X.~Qian {\it et al.}, 
  Phys.\ Rev.\  {\bf C81}, 055209 (2010).


\bibitem{Kaskulov:2011wd}
  M.~M.~Kaskulov,
  [arXiv:1105.1993 [nucl-th]].

\bibitem{Kaskulov:2008ej}
  M.~M.~Kaskulov, K.~Gallmeister, U.~Mosel,
  Phys.\ Rev.\  {\bf C79}, 015207 (2009).

\bibitem{Kaskulov:2011hr}
  M.~M.~Kaskulov, U.~Mosel,
  [arXiv:1103.1602 [nucl-th]].



\end{thebibliography}
\end{document}